# Broadband spectroscopy and interferometry with undetected photons at strong parametric amplification

Kazuki Hashimoto[1,*], Dmitri B. Horoshko[2] and Maria V. Chekhova[1,3]

[1] Max Planck Institute for the Science of Light, Staudtstr. 2, 91058 Erlangen, Germany

[2] Université de Lille, UMR 8523 - PhLAM, 42 rue Paul Duez, 59000 Lille, France

[3] Friedrich-Alexander Universität Erlangen-Nürnberg, Staudtstr. 7/B2, 91058 Erlangen, Germany

* kazuki.hashimoto@mpl.mpg.de

**Abstract**

Nonlinear interferometry with entangled photons allows for characterizing a sample without detecting the photons interacting with it. This method enables highly sensitive optical sensing in the wavelength regions where efficient detectors are still under development. Recently, nonlinear interferometry has been applied to interferometric measurement techniques with broadband light sources, such as Fourier-transform infrared spectroscopy and infrared optical coherence tomography. However, they were demonstrated with photon pairs produced through spontaneous parametric down-conversion (SPDC) at a low parametric gain, where the average number of photons per mode is much smaller than one. The regime of high-gain SPDC offers several important advantages, such as the amplification of light after its interaction with the sample and a large number of photons per mode at the interferometer output. In this study, we demonstrate broadband spectroscopy and high-resolution optical coherence tomography with undetected photons generated via high-gain SPDC in an aperiodically poled lithium niobate crystal. To prove the principle, we demonstrate reflective Fourier-transform near-infrared spectroscopy with a spectral bandwidth of 17 THz and optical coherence tomography with an axial resolution of 11 μm.

**Introduction**

Spectroscopy and interferometry with broadband light sources - e.g., Fourier-transform spectroscopy (FTS)[1] and optical coherence tomography (OCT)[2,3] - are widely used in a variety of fields, such as chemistry, biology, medicine, and industry. They noninvasively measure broadband absorption/reflection spectra or high-resolution depth profiles of samples under study, exploiting the linear interference of a broadband light source, typically created by a Michelson interferometer. For FTS and OCT, the mid-infrared (MIR) region is important because it allows for measuring fundamental vibrational modes of molecules[1] and larger depths of samples with less influence from the dispersion and scattering[4], respectively. However, infrared measurements often suffer from low measurement sensitivity because infrared detectors tend to have higher noise and lower efficiency than visible ones. Upconversion spectroscopy[5–9] or OCT[10,11] has recently been developed to improve the sensitivity of infrared measurements by exploiting wavelength conversion processes, such as sum-frequency generation (SFG) and difference-frequency generation (DFG). In addition, electro-optic (EO) sampling[12,13] techniques have been applied to broadband infrared spectroscopy. However, those methods often require high-power MIR pulsed laser sources (e.g., fs MIR ultra-short pulsed lasers), dispersion compensation schemes, and/or other W-level lasers for wavelength conversion, which adds complexity to the system.

A quantum SU(1,1) interferometer[14,15] with a narrowband pump source enables sensitive infrared measurements with a much simpler setup. It replaces the beamsplitters of a classical interferometer with nonlinear media, such as $\chi^{(2)}$ nonlinear crystals or photonic crystal fibers, to exploit pairs of entangled photons (signal and idler) generated via spontaneous parametric down-conversion (SPDC) or spontaneous four-wave mixing (SFWM), respectively. Since the phase delays of all three photons (the pump, signal, and idler) in the interferometer contribute to the signal interferences at the output[15], we can analyze the idler absorption or reflection of a sample without directly detecting the idler photons but by detecting the signal photons. This "measurement with undetected photons" can improve the sensitivity of broadband infrared spectroscopy or OCT. Several groups have already performed spectroscopy[16–24] and OCT[25–27] with undetected photons using an SU(1,1) interferometer (or a conventional scheme of "induced coherence") based on frequency nondegenerate broadband photon pairs generated via SPDC pumped with a CW laser. Those techniques are demonstrated in the low-gain regime, where the number of photons per mode generated by SPDC is much smaller than one, and the average power to be detected is up to around 100 nW[24].

Alternatively, SU(1,1) interferometry can be performed in the high-gain regime of SPDC with an intense pulsed pump source[28]. In the high-gain regime, the number of photons per mode is much larger than one because photon pairs, initially generated via SPDC, seed photon pair generation further along the crystal - that is, they undergo optical parametric amplification (OPA). The average power of the generated bright twin beams can be µW- or mW-level[28], easily detected with detectors with moderate sensitivity (e.g., a Si power meter). In addition, the seeding effect allows for noninvasively interrogating a sample with a weak idler beam and detecting it as an amplified signal beam with sufficient power for detection. Some of us recently performed OCT with undetected photons with an SU(1,1) interferometer operating in the high-gain regime[29,30]. However, the spectral bandwidth was limited to a few THz, which resulted in the OCT depth resolution of ~30 µm (an optical path-length difference (OPD) of ~60 µm ).

In this study, we develop an SU(1,1) interferometer based on broadband bright twin beams generated from an aperiodically poled lithium niobate (APLN) crystal[31–33] strongly pumped by a 532-nm picosecond pulsed laser. The spectral bandwidth of the twin beams generated from the APLN crystal is about three times larger than in the previous high-gain experiment[29]. We use the interferometer to demonstrate broadband FTS and high-resolution OCT with undetected photons. This method exploits the seeding effect of signal and idler photons inside the crystal, leading to an average detected power of up to around 10 µW. As a proof-of-principle demonstration, we measure near-infrared (NIR) reflection spectra of an optical filter at a spectral bandwidth of 17 THz and OCT depth profiles of thin samples with an axial resolution of 11 µm.

### Results
Figure 1(a) shows the schematic of an SU(1,1) interferometer with an APLN crystal pumped by picosecond pulses. The crystal has a poling period varying in a nonlinear fashion; see Supplementary Note 1 for details. The 1-mW 532-nm picosecond laser with a repetition rate of 1 kHz is focused onto an APLN crystal to generate broadband twin beams via high-gain SPDC. After passing through the crystal, the pump and generated signal beams go to the reference arm and travel back by the same path after being reflected from a spherical mirror. The idler beam is steered into the sample arm, collimated with a lens, and reflected by a reflective sample. The pump, signal, and idler beams are combined and again focused onto the APLN crystal. The signal beam is amplified or de-amplified by the APLN crystal depending on the relative phase of the pump with respect to the sum of the signal and idler phases. The amount of amplification depends on the amplitudes of the signal and idler fields. In other words, the absorption or reflection of the idler beam interrogating a sample can be detected as an intensity modulation of the signal beam (signal interferogram). The average signal power at the interferometer output is up to approximately 10 µW, and the idler power on the sample, contributing to the signal interference, is tens of nW (Supplementary Note 2). A time-domain signal interferogram for FTS is measured with a Si power meter with motorized-stage scanning and digitized with a digitizer. A spectral interferogram for OCT is measured with a visible spectrometer with a fixed stage position. The detailed schematic for the interferometer is described in the Methods section.

We first characterize the broadband twin beams generated from the APLN crystal. Figure 1(b) shows broadband signal and idler spectra individually measured with a visible (AvaSpec-ULS3648-USB2, Avantes) and a NIR (AvaSpec-NIR512-1.7TEC, Avantes) spectrometer, respectively. We focus the radiation of a 532-nm picosecond pulsed laser with an average power of around 350 µW onto the crystal using an f=100 mm lens for this evaluation. The signal and idler spectra span from 768 nm to 824 nm (364 THz – 390 THz) and from 1470 nm to 1661 nm (181 THz - 204 THz) at -10-dB intensity level, respectively. The spectral-shape difference between the twin beams is due to the wavelength-dependent sensitivity of the two spectrometers.

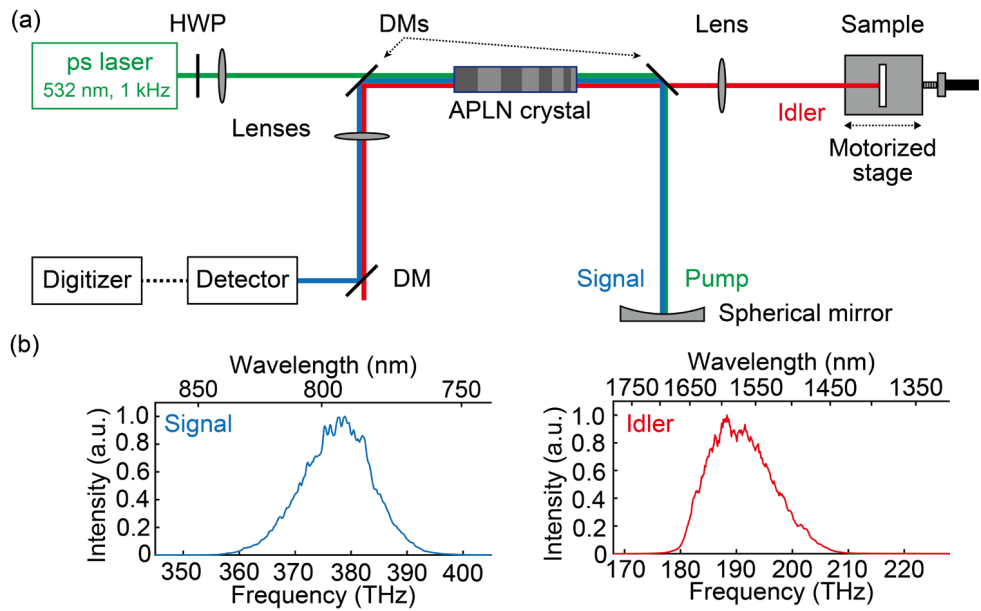

**Figure 1:** (a) Schematic of a high-gain SU(1,1) interferometer with an APLN crystal. HWP: half-wave plate, DM: dichroic mirror, APLN: aperiodically poled lithium niobate. (b) Broadband twin beams generated from an APLN crystal via nondegenerate SPDC pumped by a 532-nm picosecond pulsed laser.

Next, we measure time-domain FTS interferograms and the spectra using the SU(1,1) interferometer. The FTS interferogram measured with the interferometer is described as (see Supplementary Note 1)

$$\Delta N(\tau) = \int_0^{\omega_0} I(\Omega) R_A(-\Omega) \cos[(\omega_0 - \Omega)\tau + \rho(\Omega)] \frac{d\Omega}{2\pi}, \quad (1)$$

where $\Delta N(\tau)$ is the delay-dependent part of the total number of signal photons detected at the interferometer output. $\tau$, $\omega_0$, $\Omega$, $R_A(-\Omega)$, and $\rho(\Omega)$ denote the idler delay, the center angular frequency of SPDC light (half of the pump angular frequency), the detuning from $\omega_0$, the amplitude reflectance of a sample, and the phase acquired due to dispersive propagation in the crystal and sample, respectively. $I(\Omega)$ is proportional to $J(\Omega) = \frac{1}{2} M^4(\Omega) \sinh^2[2r(\Omega)]$, where $M(\Omega)$ and $r(\Omega)$ denote the frequency-dependent loss inside the crystal and the squeezing parameter, assuming the parametric gains of the first and second crystals are the same. Integration in Equation (1) is over the positive values of $\Omega$ corresponding to the signal band. The negative values of $\Omega$ correspond to the idler band. The phase, $\rho(\Omega)$, becomes a quadratic function due to the aperiodic poling of the crystal. The FTS spectrum is obtained by Fourier transforming Equation (1). The resultant complex spectrum is written as

$$F(\omega_0 - \Omega) = \frac{1}{2} I(\Omega) R_A(-\Omega) e^{-i\rho(\Omega)}. \quad (2)$$

where $\Omega$ is limited to the signal band $[0, \omega_0]$. Equation (2) shows that the spectral intensity, $|F(\omega_0 - \Omega)|$, is linear in the amplitude reflectance, $|R_A(-\Omega)|$. The FTS interferogram and the spectrum are discussed theoretically in detail in Supplementary Note 1.

The upper panel in Figure 2(a) shows an averaged interferogram with radiofrequency (RF) band-pass filtering. We observe sinusoidal waveforms in the centerburst of the interferogram. The dotted green line in Figure 2(b) presents the idler spectrum obtained by Fourier transforming the interferogram. The spectral bandwidth of the spectrum evaluated at -10-dB intensity level is 17 THz (135 nm). We also measure an FTS interferogram of a 1.55-µm optical band-pass filter used in the reflection geometry as a sample (lower panel in Figure 2(a)). The decay appears on one side of the interferogram because of the wavelength-dependent reflectance of the sample. The solid red line in Figure 2(b) shows the FTS spectrum of the filter. The large dip in the spectrum corresponds to the passband of the 1.55-µm band-pass filter (192 THz (1564 nm) - 196 THz (1533 nm)). The (intensity) reflectance spectrum of the filter is obtained by squaring the amplitude reflectance spectrum, derived by dividing the filter spectrum by the mirror spectrum. Figure 2(c) shows the comparison of the reflectance spectra of the filter obtained with the SU(1,1) interferometer and an optical spectrum analyzer (OSA) (AQ6374, Yokogawa). The OSA spectra (of a mirror and a filter) are measured at 2 nm spectral resolution and convoluted with a 450-GHz-width (3.6 nm at 1550 nm) sinc function to match the instrumental spectral resolution between the two spectrometers. The slight mismatch (by ~3 nm) of the cut-off wavelengths in the SU(1,1) interferometer case is probably because of the OPD mismatch between the idler beam and the Helium-Neon (HeNe) laser for the OPD axis calibration[1].

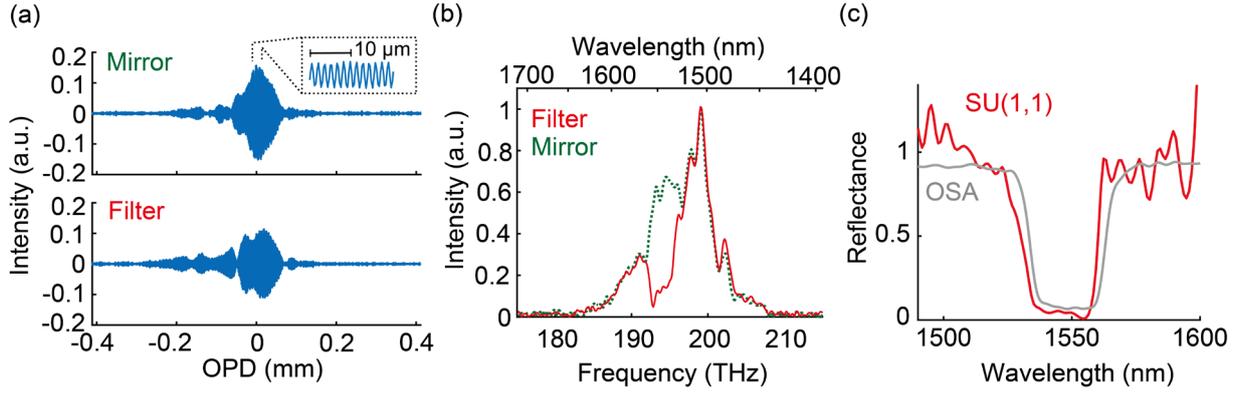

**Figure 2:** (a) Time-domain FTS interferograms of a mirror (top) and a 1.55-μm band-pass filter (bottom) used as a sample. The inset in the upper panel shows the enlarged view of the interferogram. OPD: optical path-length difference. (b) FTS reflection spectra of the mirror (dotted green) and the band-pass filter (solid red). (c) Idler (intensity) reflectance spectra of the band-pass filter measured with the SU(1,1) interferometer and an OSA.

Next, we demonstrate spectral-domain OCT using the SU(1,1) interferometer. The OCT depth profile of the sample under study is obtained by inverse Fourier transforming the spectral interferogram. We assume the sample introduces the idler delay $\tau = 2\Delta z/c$, where $2\Delta z$ and $c$ denote the OPD and the speed of light, respectively. The spectral interferogram is written as

$$S_{AC}(\omega_0 + \Omega) = J(\Omega)R_A(-\Omega)\cos[(\omega_0 - \Omega)\tau + \rho(\Omega)]. \qquad (3)$$

The interferogram is highly chirped in frequency due to the quadratic phase, $\rho(\Omega)$, which should be compensated for numerically using the Hilbert transform before applying the inverse Fourier transform. The resultant OCT peak is described as

$$i_{OCT}(z) = \left| \int_{-\infty}^{+\infty} \frac{1}{2} J(\Omega)R_A(-\Omega) e^{-i2\Omega\left(\frac{z-\Delta z}{c}\right)} \frac{d\Omega}{2\pi} \right|, \qquad (4)$$

where $z$ denotes optical depth assumed to be positive. Therefore, the waveform of the OCT peak located at $\Delta z$ is described as the inverse Fourier transform of $\frac{1}{2}J(\Omega)R_A(-\Omega)$, and thus, the linewidth is dominated by the bandwidth of the spectral interferogram. The spectral-domain OCT is discussed theoretically in detail in Supplementary Note 1.

Figure 3(a) shows the spectral interference that appeared in the measured signal spectrum when a mirror was used as a reflective sample. The frequency chirp in raw interferograms is numerically corrected by exploiting the Hilbert transform before applying the inverse Fourier transform. The correction procedures are shown in the Methods section. Figure 3(b) shows the OCT depth profiles obtained by displacing the mirror in discrete 20-μm steps, with the depth shown in the X-axis being half of the OPD between the sample and reference arms. The peak position appropriately changes following the mirror-position changes. The axial resolution is 11 μm, which is defined by the full width at half maximum (FWHM) of the peak at a depth of 50 μm. Thus, this system can measure sub-10-μm thickness materials, accounting for the typical group index of materials (>1.3). The OCT intensity gradually decreases as its depth increases because of the visible spectrometer's spectral resolution (1.4 nm, 0.640 THz at 370 THz).

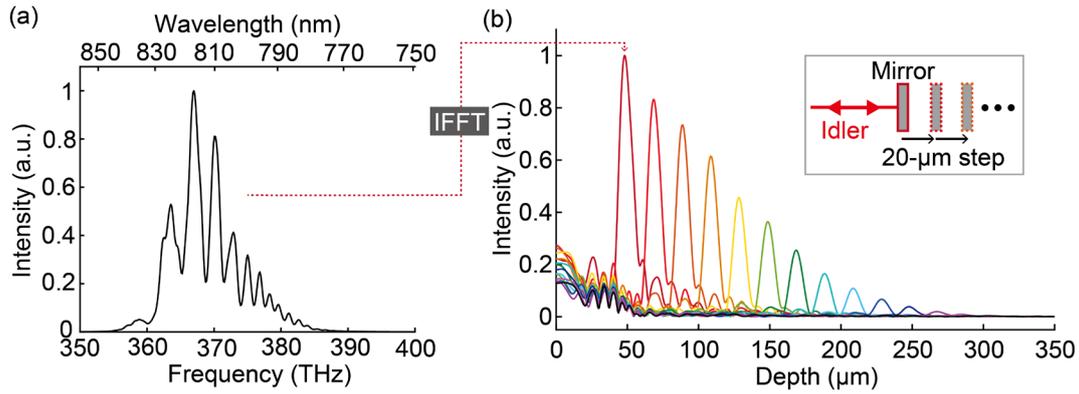

**Figure 3:** (a) A signal spectrum measured at the interferometer output. IFFT: inverse fast-Fourier transform. (b) The depth profiles of the mirror in the sample arm for the position varied in 20-μm steps. The X-axis (depth) is half of the OPD between the reference and sample arms. The OCT peaks are obtained by inverse Fourier transforming the spectral interferograms after numerical processing. The sensitivity roll-off is due to the spectral resolution of the visible spectrometer. The axial resolution, determined by the peak width at 50-μm depth, is 11 μm.

Further, we estimate the thicknesses of a lithium niobate thin film and a microscope cover glass using the SU(1,1) interferometer. Figure 4 shows the OCT depth profiles of the samples placed in the sample arm. The double peaks appearing in the profiles originate from the idler-beam reflections from the front and back surfaces of the samples. The amount of dispersion used for the numerical chirp compensation of the spectral interferograms is the same as in Figure 3. The intensity drop of the peak corresponding to the back face of the cover glass case is due to the sensitivity roll-off discussed in Figure 3(b). The optical depths of the lithium niobate thin film and cover glass evaluated by the double-peak distances are 15 μm and 156 μm, respectively. Considering the group index of the samples (lithium niobate[34]: 2.2, cover glass[35]: 1.5), we estimate the thicknesses of the lithium niobate thin film and the cover glass to be 7 μm and 102 μm, respectively, which agree well with the reference measurements shown in Supplementary Note 3 (7 μm and 104 μm, respectively). The spectral-domain OCT for the thickness measurement of a sample is also discussed theoretically in Supplementary Note 1.

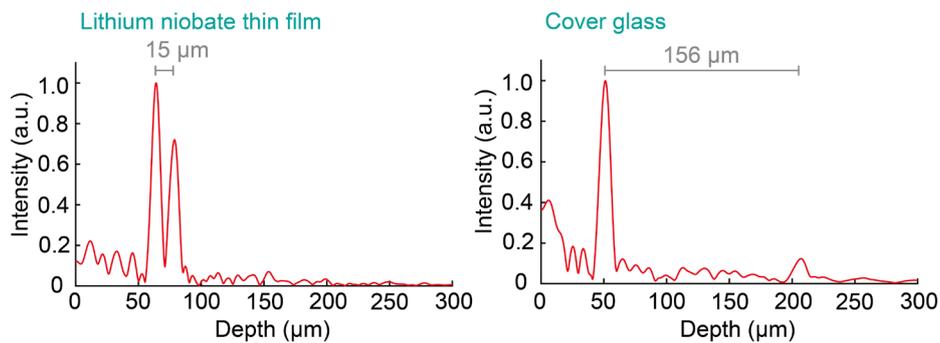

**Figure 4:** OCT depth profiles of a lithium niobate thin film and a cover glass. The double peaks correspond to the idler-beam reflections from the front and back surfaces of the samples.

## Discussion

Compared to spectroscopy with undetected photons in the low-gain regime[24], this method improved the average power by two orders of magnitude with a similar spectral bandwidth of ~20 THz. Due to the use of an aperiodically poled (chirped) crystal, this method also increased the spectral bandwidth by a factor of three compared to the previously demonstrated OCT with undetected photons in the high-gain regime[29]. Our system can be improved further by modifying the setup. First, broadband twin beams can also be generated even with a periodically poled crystal by exploiting sweet spots in the group delay curve[36]. This method can be applied to hyperspectral imaging or 3D OCT using a 2D optical sensor (e.g., a CCD or a CMOS camera) or a 2D raster scanner. In this case, the system in the high-gain regime can use an inexpensive camera with moderate detection sensitivity because it can have a large number of photons per pixel. In addition, the measurement rate can be improved by increasing the repetition rate of the pump source (under the sufficient pump photon flux to reach the high-gain regime) and applying state-of-the-art infrared spectroscopy techniques (e.g., rapid-scan FTS[37,38] or time-stretch infrared spectroscopy[9]). Furthermore, this method can be applied to multimodal spectroscopy by exploiting a part of the pump pulses for exciting other nonlinear optical phenomena[39].

## Conclusion

In conclusion, we developed an SU(1,1) interferometer based on broadband bright twin beams generated from an APLN crystal strongly pumped by a pulsed light source. With this system, we performed broadband FTS spectroscopy of an optical filter in the reflection geometry at a spectral bandwidth of 17 THz and high-resolution OCT of thin transparent samples with an axial resolution of 11 μm. We also found the FTS-spectral and the OCT-peak intensity measured with a high-gain SU(1,1) interferometer were linear in the idler amplitude reflectance, while the interference visibility was nonlinear[29]. Our system allows for spectroscopy and interferometry with undetected photons at a high average detected power of up to around 10 μW (energy per pulse: 10 nJ), which can be detected with photodetectors with moderate sensitivity. Meanwhile, the idler pulse energy on the sample (contributing to the signal interference) is expected to be tens of pJ, which is well below the typical photodamage threshold for biological samples (~1 nJ[40] but varying for different samples and the laser parameters). Our system can be beneficial for a wide range of applications that use broadband spectroscopy and high-resolution OCT.

## Methods

### SU(1,1) interferometry

A 532-nm picosecond pulsed laser with a repetition rate of 1 kHz (PL2210A-1K-SH/TH, Ekspla) is used as a pump source for the SU(1,1) interferometer. The coherence length of the pump laser evaluated by measuring the linear interferogram is 15 ps. The average pump power is around 1 mW (pulse energy of around 1μJ), and the pump polarization is adjusted with an HWP. The pump beam is focused onto a 5-mm-length type-0 APLN crystal (Gooch & Housego) using an f=200 mm lens. The poling period of the APLN crystal varies from 7.35 μm to 8.76 μm along the crystal. The inverse grating vector (defined by $2\pi/\Lambda$, where $\Lambda$ denotes the poling period) varies along the crystal as the squared hyperbolic function as described in previous works[32,33]. The high-gain SPDC process inside the crystal, pumped by the 532-nm pulsed laser, generates broadband bright twin beams (signal: ~0.8 μm, idler: ~1.6 μm). The pump, signal, and idler beams are spatially and spectrally separated with a DM with a cut-off wavelength of 980 nm (Figure 1(a)). The pump and signal beams go to the reference arm and travel back to the same path after being reflected by an f=100 mm spherical mirror. The idler beam goes to the sample arm and is collimated with an f=100-mm achromatic lens. The collimated idler beam is reflected back with a reflective sample and travels back to the same path. The sample holder is placed on a motorized stage for scanning the OPD between the reference and sample arms. The three beams are again focused onto the APLN crystal. The signal beam at the crystal output is reflected by a DM with a cut-off wavelength of 650 nm, collimated with an f=100-mm lens, and detected with a detector. A 550-nm long-pass filter and another DM (HR: 600 - 900 nm, HT: 1300 - 2200 nm) are installed in the signal beam path to filter out the pump and idler beams.

**Detection**

The time-domain FTS interferogram is detected with a Si power meter (S130VC, Thorlabs) with RF 15-Hz low-pass filtering and digitized with an 8-bit digitizer (USB-5133, National Instruments) with a sampling rate of 1.53 kSamples/s. The interferogram is recorded by moving the motorized stage in the sample arm with a scan velocity of 6 μm/s. The maximum OPD of the interferogram is 0.8 mm, corresponding to the stage displacement of 0.4 mm. The OPD of the interferogram is calibrated with a simultaneously measured HeNe interferogram. The HeNe interferogram is measured with a conventional Michelson interferometer whose scanning mirror is on the same motorized stage as the SU(1,1) interferometer. For the spectral-domain OCT measurements, the signal beam is coupled into a 400-μm-core multi-mode fiber and detected with a visible grating-based spectrometer (AvaSpec-ULS3648-USB2, Avantes) with a spectral resolution of 1.4 nm (0.640 THz at 370 THz). The power coupled into the multi-mode fiber is attenuated with neutral density filters. The spectral interferogram is measured at a fixed motorized-stage position.

**Data analysis**

The FTS and the HeNe interferograms are numerically RF band-pass filtered. The filtered FTS interferogram is resampled at the zero-crossing points of the HeNe interferogram. The corrected FTS interferogram is coherently averaged, zero-filled, and Fourier transformed to obtain the FTS (idler) spectrum. For OCT, the measured signal spectrum is divided by the background spectrum measured at a sufficiently large OPD, whose fringes are not resolvable with the visible spectrometer. The DC offset of the resultant interferogram is numerically subtracted. The dispersion appearing in the interferogram (mainly due to the aperiodic poling of the APLN crystal) is numerically corrected by extracting the complex waveform with the Hilbert transform and adding a group delay dispersion of -3500 $fs^2$ to the spectral phase[26,41]. The corrected spectral interferogram is triangular-apodized, zero-filled, and inverse Fourier transformed to obtain the OCT depth profile.

**Acknowledgments**

K. H. acknowledges the financial support by JSPS (Overseas Research Fellowships). D. B. H. acknowledges the support of Agence Nationale de la Recherche (France) via grant ANR-19-QUAN-0001 and of Franco-Bavarian University cooperation center via grant FK-09-2023. M. V. C. acknowledges the support by QuantERA II Programme (project SPARQL) that has received funding from the European Union's Horizon 2020 research and innovation programme under Grant Agreement No 101017733, with the funding organization Deutsche Forschungsgemeinschaft.

**Conflict of interest**

The authors declare no competing interests.

**Data availability**

The data that support the findings of this study are available from the corresponding author upon reasonable request.

# Supplementary Information: Broadband spectroscopy and interferometry with undetected photons at strong parametric amplification

Kazuki Hashimoto[1,*], Dmitri B. Horoshko[2] and Maria V. Chekhova[1,3]

[1] Max Planck Institute for the Science of Light, Staudtstr. 2, 91058 Erlangen, Germany

[2] Université de Lille, UMR 8523 - PhLAM, 42 rue Paul Duez, 59000 Lille, France

[3] Friedrich-Alexander Universität Erlangen-Nürnberg, Staudtstr. 7/B2, 91058 Erlangen, Germany

* kazuki.hashimoto@mpl.mpg.de

**Supplementary Note 1: Theoretical descriptions**

We consider a nonlinear interferometer illustrated in Supplementary Figure 1[1] to theoretically describe Fourier-transform spectroscopy (FTS) and optical coherence tomography (OCT) with undetected photons. The interferometer is the "unfolded" version of the Michelson nonlinear interferometer used in the experiments.

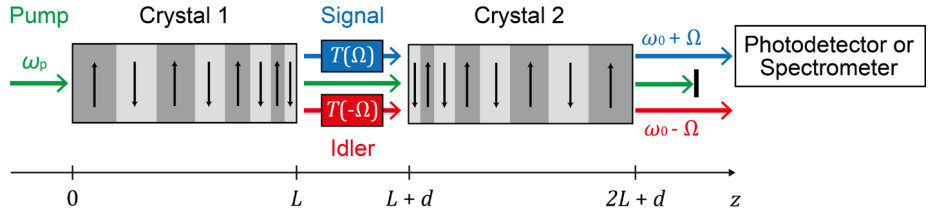

**Supplementary Figure 1**: Nonlinear interferometer. $\omega_\mathrm{p}$: pump angular frequency, $\omega_0 = \omega_\mathrm{p}/2$: center frequency of spontaneous parametric down-conversion (SPDC) light, $\Omega$: detuning from $\omega_0$, $T(\Omega)$: complex transmittance, $d$: distance between the crystals, $L$: crystal length, $z$: position.

**Bogoliubov transformation and the spectrum**

The field evolution in each crystal is given by the Bogoliubov transformations

$$a(\Omega, L) = U_1(\Omega) a(\Omega, 0) + V_1(\Omega) a^\dagger(-\Omega, 0), \quad (S1)$$

$$a(\Omega, z_\mathrm{out}) = U_2(\Omega) a(\Omega, L+d) + V_2(\Omega) a^\dagger(-\Omega, L+d), \quad (S2)$$

where $\Omega$, $d$, $L$, $z_\mathrm{out}$, and $a(\Omega, z)$ denote the detuning from the center angular frequency of spontaneous parametric down-conversion (SPDC) light $\omega_0$, the distance between the crystals, the length of each crystal, the position of the second crystal output $z_\mathrm{out} = 2L + d$, and the annihilation operator for the photon of frequency $\omega_0 + \Omega$ at position $z$, respectively. The functions $U_n(\Omega)$ and $V_n(\Omega)$, where $n$ denotes the crystal number, satisfy the relations $|U_n(\Omega)|^2 - |V_n(\Omega)|^2 = 1$ and $U_n(\Omega)/V_n(\Omega) = U_n(-\Omega)/V_n(-\Omega)$ required by the unitarity of the field transformations. We denote the phase acquired between the crystals by the field component at the frequency $\omega_0 + \Omega$ as $\phi(\Omega)$ and define the complex transmittance between the crystals $T(\Omega) = T_A(\Omega) e^{i\phi(\Omega)}$, where $T_A(\Omega)$ is real (but not necessarily positive). We note the formulas for the Michelson interferometer in the main text are obtained by replacing $T_A(\Omega)$ with $R_A(\Omega)$ in the formulas for the "unfolded" interferometer. The field transformation between the crystals is written as

$$a(\Omega, L + d) = T(\Omega)a(\Omega, L) + R(\Omega)a_{\text{vac}}(\Omega), \tag{S3}$$

where $R(\Omega)$ is the corresponding complex "reflectance", making the entire transformation unitary thanks to the relation $|T(\Omega)|^2 + |R(\Omega)|^2 = 1$, and $a_{\text{vac}}(\Omega)$ is a vacuum field added due to absorption in the sample. Substituting Equations (S1) and (S3) into (S2), we obtain the total input-output field transformation

$$a(\Omega, z_{\text{out}}) = U_{\text{int}}(\Omega)a(\Omega, 0) + V_{\text{int}}(\Omega)a^\dagger(-\Omega, 0) + U_2(\Omega)R(\Omega)a_{\text{vac}}(\Omega) + V_2(\Omega)R^*(-\Omega)a^\dagger_{\text{vac}}(-\Omega), \tag{S4}$$

where

$$U_{\text{int}}(\Omega) = U_2(\Omega)U_1(\Omega)T(\Omega) + V_2(\Omega)V_1^*(-\Omega)T^*(-\Omega), \tag{S5}$$

$$V_{\text{int}}(\Omega) = U_2(\Omega)V_1(\Omega)T(\Omega) + V_2(\Omega)U_1^*(-\Omega)T^*(-\Omega), \tag{S6}$$

are the Bogoliubov coefficients for the entire interferometer. The power spectrum of the signal wave ($\Omega > 0$) at the interferometer output is defined by the relation $\langle a^\dagger(\Omega, z_{\text{out}})a(\Omega', z_{\text{out}})\rangle = 2\pi S(\omega_0 + \Omega)\delta(\Omega - \Omega')$, where $\delta$ denotes the Dirac delta function. From Equation (S4) and the equal-space commutation relations $[a(\Omega, z), a^\dagger(\Omega', z)] = 2\pi\delta(\Omega - \Omega')$, we obtain

$$S(\omega_0 + \Omega) = |V_{\text{int}}(\Omega)|^2 + |V_2(\Omega)R(-\Omega)|^2. \tag{S7}$$

**Parametrization**

Each Bogoliubov transformation can be parametrized by four real functions of the frequency detuning[2]:

$$r_n(\Omega) = \ln(|U_n(\Omega)| + |V_n(\Omega)|), \tag{S8}$$

$$\psi_n^0(\Omega) = \frac{1}{2}\arg[U_n^{-1}(\Omega)V_n(\Omega)], \tag{S9}$$

$$\psi_n^L(\Omega) = \frac{1}{2}\arg[U_n(\Omega)V_n(-\Omega)], \tag{S10}$$

$$\kappa_n(\Omega) = \frac{1}{2}\arg[U_n(\Omega)U_n^{-1}(-\Omega)], \tag{S11}$$

where $r_n(\Omega)$ is known as the squeezing parameter at frequency $\omega_0 + \Omega$, while $\psi_n^{0(L)}(\Omega)$ is the input (output) squeezing angle. The first three parameters are even functions of $\Omega$, while the fourth one is odd. Alternatively, we can write

$$U_n(\Omega) = e^{i[\psi_n^L(\Omega) - \psi_n^0(\Omega) + \kappa_n(\Omega)]}\cosh r_n(\Omega), \tag{S12}$$

$$V_n(\Omega) = e^{i[\psi_n^L(\Omega) + \psi_n^0(\Omega) + \kappa_n(\Omega)]}\sinh r_n(\Omega). \tag{S13}$$

**Aperiodically poled crystal and the quantum Rosenbluth formula**

We consider a crystal of lithium niobate doped by 5% magnesium oxide of length $L = 5$ mm illuminated by a monochromatic pump beam at a wavelength $\lambda_p = 532$ nm polarized as an extraordinary wave of the crystal. In the process of SPDC, two extraordinary waves (signal and idler) collinearly propagate with the pump. The crystal is aperiodically poled for quasi-phase-matching (QPM) of the idler wave between $\lambda_i^{\text{low}} = 1.42$ μm and $\lambda_i^{\text{high}} = 2.13$ μm, corresponding to the signal wavelengths between $\lambda_s^{\text{high}} = 0.85$ μm and $\lambda_s^{\text{low}} = 0.71$ μm.

The refractive index of the extraordinary wave in MgO-doped lithium niobate, $n_e(\omega)$, is determined by the Sellmeier equations at a temperature of 24.5 C°[3]. We calculate the dispersion law $k(\Omega) = \omega n_e(\omega)/c$, where $\omega$ and $c$ denote the angular frequency and the speed of light, and define the phase mismatch

$$\Delta(\Omega) = k_p - k(\Omega) - k(-\Omega), \tag{S14}$$

where $k_p$ denotes the wave vector of the pump wave in the crystal. This function can be approximated by a quadratic dependence

$$\Delta_q(\Omega) = -\alpha \left(\frac{\Omega}{\omega_0}\right)^2 + \beta, \tag{S15}$$

where $\alpha = 735$ rad/mm and $\beta = 901$ rad/mm are two constant values[4].

The QPM of the target spectral ranges is attained by the aperiodic poling of the crystal. The inverse grating vector, $K = 2\pi/\Lambda$, $\Lambda$ being the poling period, varies from $K_0 = \Delta(\Omega^0) = 720$ rad/mm to $K_L = \Delta(\Omega^L) = 855$ rad/mm, where $\Omega^0 = 2\pi c/\lambda_s^{low} - \omega_0 = 2\pi \times 140$ THz and $\Omega^L = 2\pi c/\lambda_s^{high} - \omega_0 = 2\pi \times 70$ THz are the signal detunings quasi-phase-matched at the input and output of the crystal surfaces, respectively. The poling profile of $K(z)$ between these two values is chosen as a quadratic-hyperbolic profile[1,4,5]

$$K(z) = -\left(\frac{\Omega^0 \Omega^L}{\omega_0}\right)^2 \frac{\alpha}{[\Omega^L - (\Omega^L - \Omega^0) z/L]^2} + \beta. \tag{S16}$$

The field transformation in an aperiodically poled crystal with the poling profile $K(z)$ is described approximately by the Bogoliubov coefficients, given by the "quantum Rosenbluth formula"[4]:

$$U(\Omega) = e^{\pi \nu(\Omega)} e^{i(k(\Omega)-k_0)L}, \tag{S17}$$

$$V(\Omega) = \sqrt{e^{2\pi \nu(\Omega)} - 1}\, e^{-2i\xi(\Omega)+i(k(\Omega)-k_0)L+i\varphi_A}, \tag{S18}$$

where $\xi(\Omega)$ is the phase accumulated due to the crystal dispersion, $\varphi_A$ is the gain-dependent phase, and $k_0$ is the wave vector at the center frequency of SPDC light. We introduce the frequency-dependent Rosenbluth parameter $\nu(\Omega)$ as

$$\nu(\Omega) = \frac{\nu_0}{2} \frac{\Omega^L \Omega^0 (\Omega^L + \Omega^0)}{|\Omega|^3}, \tag{S19}$$

defined via the "average" Rosenbluth parameter $\nu_0 = |\gamma|^2 L/|K_L - K_0|$, where $\gamma$ is the coupling constant for the three-wave mixing process, proportional to the pump amplitude.

**Spectral filtering**

We build a model for spectral filtering inside the crystal because the experimentally measured signal and idler spectra have narrower bandwidths than the ones supposed by the crystal design (~30 THz against ~70 THz), which is caused by some blocking mechanism and which we ascribe to scattering on the inhomogeneities appearing in the aperiodically poled lithium niobate (APLN) crystal because of fabrication imperfections. In the model, the field evolution in the APLN crystal is described by the Bogoliubov transformation, Equations (S1) and (S2), and a subsequent filtering transformation with the amplitude transmittance $M(\Omega): a(\Omega, L) \to M(\Omega)a(\Omega, L) + \widetilde{M}(\Omega)b_{\text{vac}}(\Omega)$, where $\widetilde{M}(\Omega)$ is the corresponding reflectance, satisfying $|M(\Omega)|^2 + |\widetilde{M}(\Omega)|^2 = 1$, and $b_{\text{vac}}(\Omega)$ is the vacuum field. $M(\Omega)$ is assumed to be a real even function. After the first passage, the field operator is

$$a(\Omega, L) = M(\Omega)U_1(\Omega)a(\Omega, 0) + M(\Omega)V_1(\Omega)a^\dagger(-\Omega, 0) + \widetilde{M}(\Omega)b_{\text{vac}}(\Omega), \tag{S20}$$

and its spectrum is $S_1(\omega_0 + \Omega) = |M(\Omega)V_1(\Omega)|^2$. We choose a super-Gaussian shape of the filter

$$M(\Omega) = M_0 e^{-\frac{(\Omega-\Omega_m)^4}{4\sigma_m^4}}, \tag{S21}$$

where $M_0 = 5 \times 10^{-2}, \omega_0 + \Omega_m = 2\pi \times 400$ THz, $\sigma_m = 2\pi \times 19$ THz. We calculate the twin beam spectra after the first passage using Equations (S18) – (S21) assuming $\nu_0 = 1$ (Supplementary Figure 2).

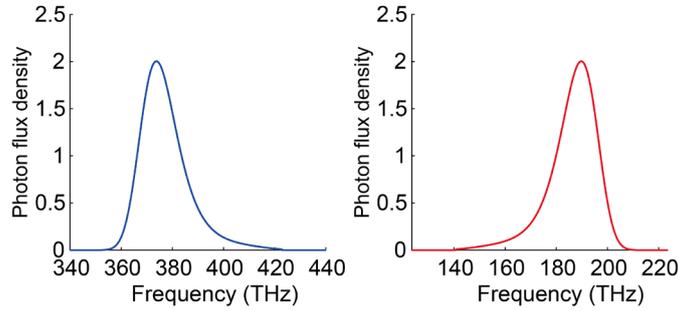

**Supplementary Figure 2:** Simulated signal (left) and idler (right) spectra after the first passage.

**Spectra after the second passage**

The field operator after the second passage is described as:

$$\begin{aligned}
a(\Omega, z_{\text{out}}) &= \widetilde{U}_{\text{int}}(\Omega)a(\Omega, 0) + \widetilde{V}_{\text{int}}(\Omega)a^\dagger(-\Omega, 0) \\
&+ M(\Omega)U_2(\Omega)R(\Omega)a_{\text{vac}}(\Omega) + M(\Omega)V_2(\Omega)R^*(-\Omega)a^\dagger_{\text{vac}}(-\Omega) \\
&+ M(\Omega)\widetilde{M}(\Omega)U_2(\Omega)T(\Omega)b_{\text{vac}}(\Omega) + M(\Omega)\widetilde{M}^*(\Omega)V_2(\Omega)T^*(-\Omega)b^\dagger_{\text{vac}}(\Omega) \\
&+ \widetilde{M}(\Omega)c_{\text{vac}}(\Omega),
\end{aligned} \tag{S22}$$

where $\widetilde{U}_{\text{int}}(\Omega) = M^2(\Omega)U_{\text{int}}(\Omega), \widetilde{V}_{\text{int}}(\Omega) = M^2(\Omega)V_{\text{int}}(\Omega)$, and $c_{\text{vac}}(\Omega)$ is the vacuum field. The signal ($\Omega > 0$) and idler ($\Omega < 0$) spectra are given by

$$S_2(\omega_0 + \Omega) = |\widetilde{V}_{\text{int}}(\Omega)|^2 + M^2(\Omega)|V_2(\Omega)R^*(-\Omega)|^2 + M^2(\Omega)\widetilde{M}^2(\Omega)|V_2(\Omega)T^*(-\Omega)|^2, \tag{S23}$$

where only the first term depends on the phase of $T(\Omega)$. The phase acquired by the field between the crystals is written as

$$\phi(\Omega) = \tau_0 \Omega + \tau(\omega_0 + \Omega)\theta(-\Omega) + \phi_0(\Omega)\theta(-\Omega), \tag{S24}$$

where $\tau_0$ is the initial delay time for both the signal and idler beams, $\tau$ is the additional idler delay, $\phi_0(\Omega)$ is the phase acquired by the idler field in the sample, and $\theta(\Omega)$ is the step function, which is equal to one for $\Omega > 0$ and zero for $\Omega < 0$. We recall that at any point the phase of the sideband operator $a(\Omega, z)$ is defined relatively to the hypothetic central-frequency wave at $\omega_0$. As a consequence, the term $\tau\omega_0$ describes the phase acquired by this central-frequency wave in the delay line. Assuming that the pump is undepleted and therefore $|U_2| = |U_1|, |V_2| = |V_1|, r_2 = r_1 = r$ in the regime of a high blocking ratio $M_0 \ll 1$, neglecting the fourth-order in $M_0$ term as compared to the quadratic one in the delay-independent part of the spectrum and using $|R(-\Omega)|^2 + |T(-\Omega)|^2 = 1$, we rewrite Equation (S23) for $\Omega > 0$ as

$$S_2(\omega_0 + \Omega) = J(\Omega)\left(\frac{1 + |T_A(-\Omega)|^2}{2} + T_A(-\Omega)\cos[(\omega_0 - \Omega)\tau + \rho(\Omega)]\right) + M^2(\Omega)\sinh^2[r(\Omega)], \quad (S25)$$

where $J(\Omega) = \frac{1}{2}M^4(\Omega)\sinh^2[2r(\Omega)]$, $\rho(\Omega) = \phi_0(-\Omega) + 2\psi_1^L(\Omega) - 2\psi_2^0(\Omega)$, and we have assumed $T(\Omega) = 1$ for $\Omega > 0$. Using Equations (S17) and (S18), we obtain (up to a constant phase)[1]

$$\psi_1^L(\Omega) = -\psi_2^0(\Omega) = -\frac{\alpha L}{2\omega_0^2}\frac{\Omega^0(\Omega - \Omega^L)^2}{\Omega^L - \Omega^0}, \quad (S26)$$

wherefrom, assuming $\phi_0(\Omega) = 0$, we obtain $\rho(\Omega) = 4\psi_1^L(\Omega)$. The signal power after the second passage is

$$P_{2,\text{signal}} = \int_{\omega_0}^{\omega_\text{p}} \hbar\omega S_2(\omega)\frac{d\omega}{2\pi}. \quad (S27)$$

For $\nu_0 = 1$ and $T(-\Omega) = 1$, the signal power is 38 μW. In the case of $J(\Omega) \gg M^2(\Omega)\sinh^2[r(\Omega)]$, the visibility of the spectrum $S_2(\omega_0 + \Omega)$ is approximated as

$$Visibility \approx \frac{2|T_A(-\Omega)|}{1 + |T_A(-\Omega)|^2}. \quad (S28)$$

The visibility is nonlinear in the amplitude transmittance.

**FTS with undetected photons**

Next, we theoretically describe FTS with undetected photons. We regard the pump as a quasi-monochromatic wave switched on at time $t = 0$ and off at $t = \tau_\text{p}$. Then, the total number of photons detected by the signal photodetector is

$$N(\tau) = \eta \int_0^{\tau_\text{p}} \langle E^{(-)}(t, z_\text{out})E^{(+)}(t, z_\text{out})\rangle dt, \quad (S29)$$

where $\eta$ denotes the quantum efficiency of the photodetector and the field in photon-flux units is

$$E^{(+)}(t, z) = \int a(\Omega, z)e^{i[k_0 z - (\omega_0 + \Omega)t]}\frac{d\Omega}{2\pi}. \quad (S30)$$

Substituting Equation (S22) into Equations (S29) and (S30), we obtain

$$N(\tau) = \eta\tau_\text{p} \int_0^{\omega_0} S_2(\omega_0 + \Omega)\frac{d\Omega}{2\pi}$$
$$= N_0 + \int_0^{\omega_0} I(\Omega)T_A(-\Omega)\cos[(\omega_0 - \Omega)\tau + \rho(\Omega)]\frac{d\Omega}{2\pi}, \quad (S31)$$

where $I(\Omega) = \eta\tau_\text{p}J(\Omega)$ and $N_0$ is a term independent of $\tau$ (the DC component of the time-domain waveform). The FTS interferogram is obtained by subtracting the constant term, $\Delta N(\tau) = N(\tau) - N_0$, thus

$$\Delta N(\tau) = \int_0^{\omega_0} I(\Omega) T_A(-\Omega) \cos[(\omega_0 - \Omega)\tau + \rho(\Omega)] \frac{d\Omega}{2\pi}. \tag{S32}$$

Supplementary Figure 3 shows the simulated interferograms with and without a 194-THz band-stop filter with a bandwidth of 4 THz (as a sample), which corresponds to a band-pass filter in the Michelson geometry used in the experiment.

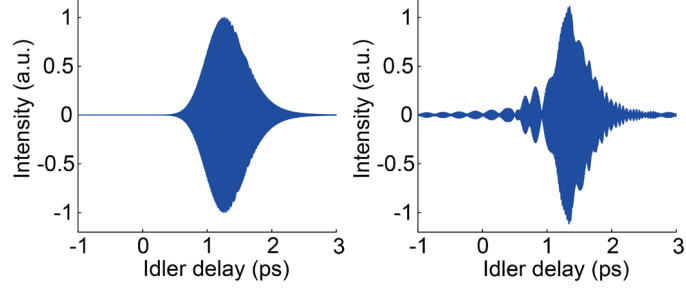

**Supplementary Figure 3:** Simulated interferograms with (right) and without (left) a 194-THz band-stop filter, whose bandwidth and amplitude transmittance (at 194 THz) are 4 THz and 0.1, respectively.

The spectrum is obtained by complex Fourier transforming the interferogram:

$$F(\omega) = \int_{-\infty}^{+\infty} \Delta N(\tau) e^{i\omega\tau} d\tau. \tag{S33}$$

If we limit the angular frequency $\omega$ to the idler band $\omega \in [0, \omega_0])$, then

$$F(\omega_0 - \Omega) = \frac{1}{2} I(\Omega) T_A(-\Omega) e^{-i\rho(\Omega)}. \tag{S34}$$

The amplitude transmittance can be obtained as the ratio

$$|T_A(-\Omega)| = \frac{|F(\omega_0 - \Omega)|}{|F_0(\omega_0 - \Omega)|}, \tag{S35}$$

where $F_0$ is measured without the sample. Therefore, the spectral intensity is linear in the amplitude transmittance (i.e., in the square root of the intensity transmittance), while the time-domain interference visibility is nonlinear[7]. This is because the FTS spectrum is obtained from only the AC part of the time-domain waveform, while the visibility considers the DC part. Supplementary Figure 4(a) shows the spectra with and without the band-stop filter obtained by Fourier transforming the interferograms shown in Supplementary Figure 3, and Supplementary Figure 4(b) shows the transmittance spectrum of the filter obtained by dividing the spectrum with the filter by the one without the filter. The transmittance at 194 THz is linear in the idler amplitude transmittance of the band-stop filter, as shown in Supplementary Figure 4(c).

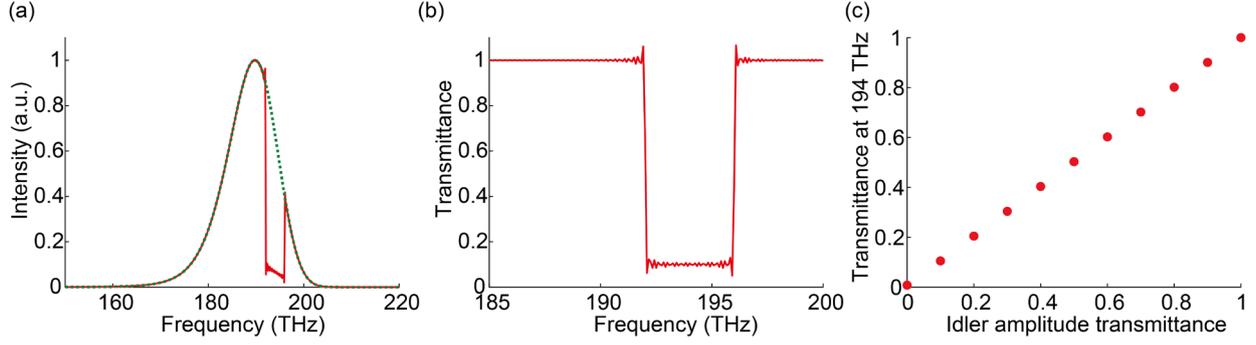

**Supplementary Figure 4:** (a) Spectra with (solid red) and without (dotted green) the filter obtained by Fourier transforming the simulated interferograms in Supplementary Figure 3. (b) The transmittance spectrum of the band-stop filter. (c) Transmittance at 194 THz against the idler amplitude transmittance of the band-stop filter.

**Spectral-domain OCT with undetected photons**

We also theoretically describe spectral-domain OCT with undetected photons. First of all, we consider measurements with a mirror in the idler path placed a distance $\Delta z$ further from the crystal than the mirror in the signal path, which is a similar situation as Figure 3. This means the idler delay is $\tau = 2\Delta z/c$ for the "unfolded" interferometer of Supplementary Figure 1. Supplementary Figure 5(a) shows the simulated spectrum for $\tau = 0.33$ ps. The spectral interferogram is chirped with respect to the frequency due to the quadratic term $\rho(\Omega)$.

Spectral-domain OCT consists of taking the inverse Fourier transform of the measured spectrum and analyzing its positive peaks. The cosine term of Equation (S25) is the spectral interferogram

$$S_{AC}(\omega_0 + \Omega) = J(\Omega)T_A(-\Omega)\cos[(\omega_0 - \Omega)\tau + \rho(\Omega)], \tag{S36}$$

which only contributes to the OCT depth profile after the inverse Fourier transform. The quadratic term $\rho(\Omega)$ can be compensated for numerically by exploiting the Hilbert transform[8,9]. The resultant complex spectral interferogram is written as

$$S_H(\omega_0 + \Omega) = \frac{1}{2}J(\Omega)T_A(-\Omega)e^{-i(\omega_0-\Omega)\tau}. \tag{S37}$$

The absolute value of the inverse Fourier transform of the complex spectral interferogram is

$$\begin{aligned}i_{OCT}(z) &= \left|\int_{-\infty}^{+\infty} S_H(\omega_0 + \Omega) e^{-i2\Omega\frac{z}{c}}\frac{d\Omega}{2\pi}\right| \\ &= \left|\int_{-\infty}^{+\infty} \frac{1}{2}J(\Omega)T_A(-\Omega) e^{-i2\Omega\left(\frac{z-\Delta z}{c}\right)}\frac{d\Omega}{2\pi}\right|.\end{aligned} \tag{S38}$$

Supplementary Figure 5(b) shows the inverse Fourier-transformed result of the spectral interferogram shown in (a), assuming the quadratic phase is removed. The OCT-peak intensity is proportional to $T_A(-\Omega)$ if $T_A(-\Omega)$ is constant in frequency, as shown in Supplementary Figure 5(c).

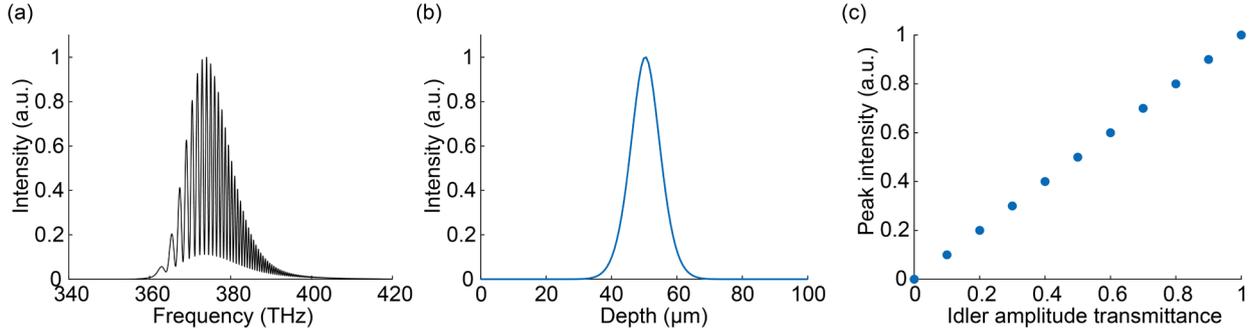

**Supplementary Figure 5:** (a) A simulated spectrum after the second passage with $T_A(-\Omega) = 1$ and $\tau = 0.33$ ps ($\Delta z = 50$ μm) (b) The OCT depth profile obtained by inverse Fourier transforming the spectral interferogram in (a), assuming the quadratic phase is removed ($\rho(\Omega) = 0$). (c) The OCT-peak intensity against $T_A(-\Omega)$, assuming $T_A(-\Omega)$ is constant in frequency.

We also consider the thickness measurement of thin samples with spectral-domain OCT. The field reflected from a thin sample with parallel surfaces at normal incidence in the low-reflectance limit is a superposition of the fields reflected from the first and the second surfaces, that is

$$E_{\text{out}}^{(+)}(t,z) = R_s E_{\text{in}}^{(+)}(t - 2\Delta z/c, z) + R_s E_{\text{in}}^{(+)}(t - 2\Delta z/c - 2d_{\text{OCT}} n_s/c, z), \tag{S39}$$

where $E_{\text{in}}^{(+)}(t,z)$, $R_s$, $\Delta z$, $d_{\text{OCT}}$, and $n_s$ denote the incident field, the reflectance of each surface, the distance from the point $z$, the sample width, and the sample group index, respectively. Substituting this equation into Equation (S30), we obtain the corresponding relation for the photon annihilation operators. For the "unfolded" interferometer of Supplementary Figure 1, the complex transmittance at $\Omega > 0$ is described as

$$T(-\Omega) = T_0 \left[ e^{i(\omega_0 - \Omega)\tau_1} + e^{i(\omega_0 - \Omega)(\tau_1 + \tau_s)} \right], \tag{S40}$$

where $T_0 = R_s$, $\tau_1 = 2\Delta z/c$ is the delay of the wave reflected from the first surface, and $\tau_s = 2d_{\text{OCT}} n_s/c$ is the round-trip time of the wave in the sample. Here, we have neglected the round-trip time of the signal arm $\tau_0$, which does not affect the interference. We write the amplitude $T_A(-\Omega)$ and phase $\phi(-\Omega)$ as

$$T_A(-\Omega) = 2T_0 \cos\left[(\omega_0 - \Omega)\frac{\tau_s}{2}\right], \tag{S41}$$

$$\phi(-\Omega) = (\omega_0 - \Omega)\left(\tau_1 + \frac{\tau_s}{2}\right). \tag{S42}$$

The signal spectrum after the second passage is given by Equation (S25) with $\tau = \tau_1 + \tau_s/2$. The complex spectral interferogram after the signal processing is written as

$$S_H(\omega_0 + \Omega) = J(\Omega) T_0 \cos\left[(\omega_0 - \Omega)\frac{\tau_s}{2}\right] e^{-i(\omega_0 - \Omega)(\tau_1 + \frac{\tau_s}{2})}, \tag{S43}$$

where $|T_A(-\Omega)|^2$ term in Equation (S25) is neglected due to the low-reflectance limit, and the phase $\rho(\Omega)$ is removed by the signal processing. The OCT depth profile is described as

$$i_{\text{OCT}}(z) = T_0 \left| \int_{-\infty}^{+\infty} \frac{1}{2} J(\Omega) \left[ e^{-i2(\omega_0 - \Omega)\frac{\Delta z}{c}} + e^{-i2(\omega_0 - \Omega)\left(\frac{\Delta z + d_{\text{OCT}} n_s}{c}\right)} \right] e^{-i2\Omega \frac{z}{c}} \frac{d\Omega}{2\pi} \right|. \tag{S44}$$

This equation has doublet peaks at $z_1 = \Delta z$ and $z_2 = \Delta z + d_{\text{OCT}} n_s$. Supplementary Figure 6 shows the simulated

OCT depth profiles of the thin samples used in the experiment (left: lithium niobate thin film, $d_{OCT} = 7$ μm and $n_s = 2.2$, right: cover glass, $d_{OCT} = 104$ μm and $n_s = 1.5$) with $T_0 = 0.1$ and $\Delta z = 50$ μm.

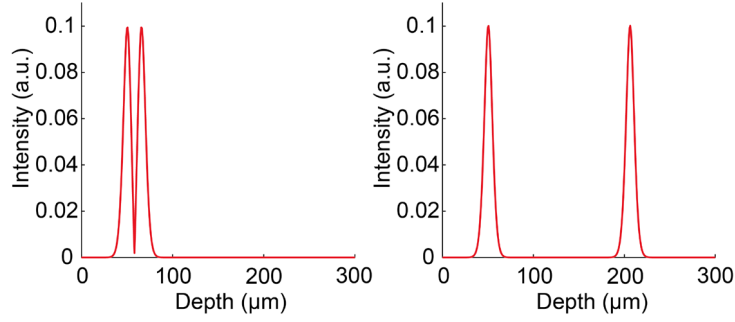

**Supplementary Figure 6:** (a) Simulated OCT depth profiles of thin samples with $d_{OCT} = 7$ μm and $n_s = 2.2$ (left) and $d_{OCT} = 104$ μm and $n_s = 1.5$ (right). We assume $\Delta z = 50$ μm, $T_0 = 0.1$, and $|T_A(-\Omega)|^2$-term in Equation (S25) can be neglected.

**Supplementary Note 2: Characterization of the system**

**Idler intensity after the first passage**

We measure the idler intensity after the first passage versus the average pump power (Supplementary Figure 7). The average idler power is monitored with a Ge power meter (S132C, Thorlabs) in the sample arm, while the average pump power is measured with a Si power meter (S130VC, Thorlabs). An iris is installed before the Ge power meter to collect the central part of the collimated idler beam, contributing to signal interference in our experiments. The idler intensity exponentially increases as the pump power increases and saturates at a pump power of over 1.1 mW. The saturation is due to the depletion of the pump beam inside the crystal[10]. We also roughly estimate the gain inside the crystal from the plot. Assuming the Rosenbluth parameter[4] is constant in frequency: $\nu_{const}$ for simplicity, we can represent the intensity as $I \propto e^{2\pi\nu_{const}} - 1$. The first seven data points in the plot are fitted with an exponential function, $y = A(e^{Bx} - 1)$, where $A = 3.05^{-17}$ and $B = 19.2$ are the fitting coefficients. The estimated $\nu_{const}$ at a pump power of 1 mW is around 3, and the idler power onto the sample contributing to the signal interference is around 10 nW. The calculations in Supplementary Note 1 show the idler power after the first passage is 5 μW at $\nu_0 = 1$, while the experiment shows 10 nW at $\nu_{const} = 3$. One of the possible reasons for the mismatch is that we collect only the central part of the collimated idler beam. To carefully evaluate the mismatch, we need to accurately measure the frequency-dependent Rosenbluth parameter by selecting the spatiotemporal modes of the SPDC light and also compare the results with the calculations[5]. Also, some differences between the model and the observations can be ascribed to a limited applicability of the model with a CW pump, developed here and implying a continuum of signal and idler frequency modes, to experiments with a pulsed pump, where a finite discrete set of frequency modes can be introduced. A modal theory of pulsed SPDC in aperiodically poled crystals will be the subject of a separate study.

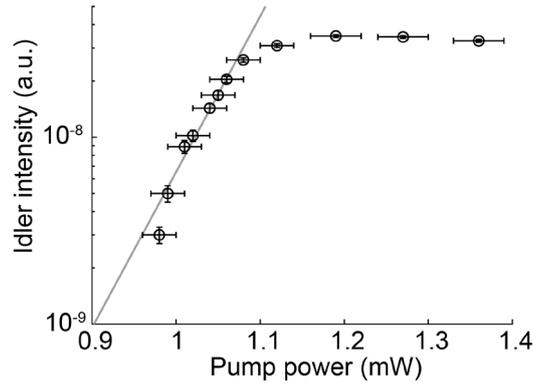

**Supplementary Figure 7:** Idler intensity after the first passage against the pump power (solid gray line: curve fitting the first seven data points with an exponential function). The fitting coefficients for the exponential function $y = A(e^{Bx} - 1)$ are $A = 3.05^{-17}$ and $B = 19.2$, respectively. The X- and Y-axis error bars represent the standard deviation of the pump power and the idler intensity, respectively.

### Spectral intensity after the second passage

We measure the signal spectral intensity after the second passage by varying the average pump power (Supplementary Figure 8 (a)). The spectra are obtained with the same setup as Figure 3 at different pump powers and recorded by changing the exposure time of the spectrometer and/or installing neutral density (ND) filters to avoid the saturation of the spectrometer. The relative spectral intensity is retrieved assuming the linearity of the spectrometer's exposure time and the ND filters' attenuation. Supplementary Figure 8 (b) represents the mean spectral intensity at the green- and red-shaded areas versus the pump power. The intensity increases as the pump power increases and finally saturates. We adjust the pump power for OCT measurements up to 1.1 mW to avoid the saturation regime. The signal power at a pump power of 1.08 mW is about 9 μW, sufficient for detecting with a photodetector with moderate sensitivity.

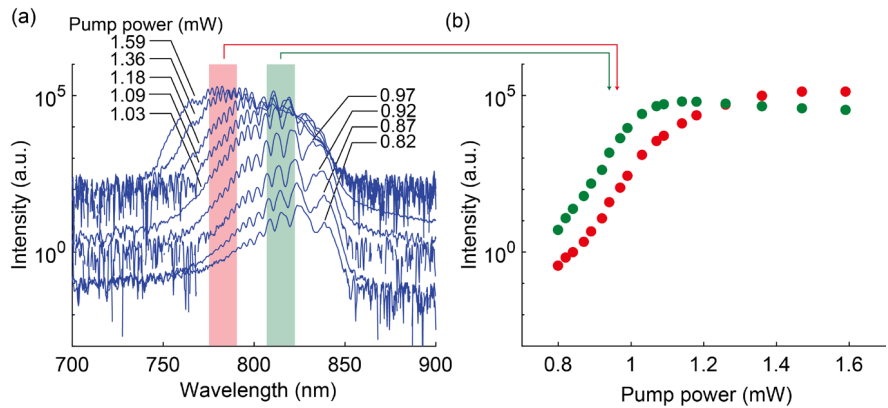

**Supplementary Figure 8:** (a) Signal spectra after the second passage measured at several average pump powers. The noise level changes because the exposure time is changed at each measurement to avoid the saturation of the spectrometer. (b) The mean spectral intensity in the green- and red-shaded areas in (a) against the pump power.

### FTS-spectral and OCT-peak intensity as functions of the idler amplitude transmittance

We experimentally evaluate FTS-spectral intensity dependence on the idler amplitude transmittance. The FTS-spectral intensity is traced by changing the amount of the idler power with the variable ND filter, which is the absorption sample in this measurement. The attenuated beam returns to the same path after being reflected by a mirror. The idler power is individually monitored with a Ge power meter to calculate the idler amplitude transmittance of the sample in the double-pass geometry. Supplementary Figure 9 (a) shows the FTS-spectral intensity (integrated from 1538 nm to 1549 nm) against idler amplitude transmittance. As expected in the simulations, the measured FTS-spectral intensity is proportional to the idler amplitude transmittance. The slight mismatch between the measured plots and the fitting curve at lower transmittance is probably due to the inaccuracy of the Ge power meter at low idler power. Using the same configuration, we also experimentally evaluate the OCT-peak intensity (located at an optical depth of 83 μm) as a function of the idler amplitude transmittance (Supplementary Figure 9 (b)). The OCT-peak intensity is also proportional to the idler amplitude transmittance.

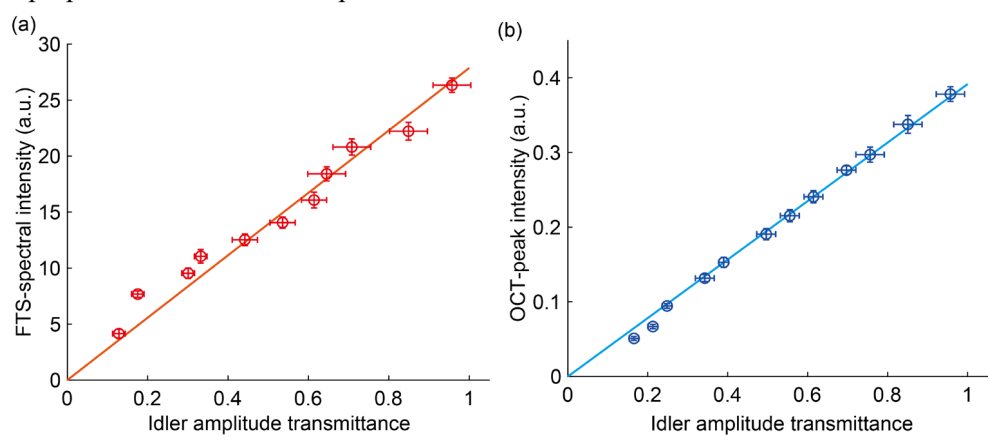

**Supplementary Figure 9:** (a) FTS-spectral intensity versus the idler amplitude transmittance of the sample (solid orange line: curve fitting with a linear function). The X- and Y-axis error bars represent the standard deviation of the idler amplitude transmittance and the FTS-spectral intensity, respectively. (b) OCT-peak intensity versus the idler amplitude transmittance of the sample (solid sky-blue line: curve fitting with a linear function). The X- and Y-axis error bars represent the standard deviation of the idler amplitude transmittance and the OCT-peak intensity, respectively.

**Supplementary Note 3: Reference measurements**

The thickness of the lithium niobate thin film and the cover glass utilized for the OCT measurements are determined with the Fabry-Pérot etalon measurements, exploiting a white light source and an OSA. The collimated white light source is directed onto the sample, and the reflected light is measured with the OSA. The etalon effect inside the samples produces interference on the spectrum, whose inverse Fourier-transformed results contain information about the optical depth between the sample surfaces. Supplementary Figure 10 shows the optical depths between the samples' front- and back-side surfaces. The peaks, with a linewidth of around 4 μm, are obtained by inverse Fourier transforming the spectral interferograms spanning from 1200 nm to 1650 nm. The optical depths of a lithium niobate thin film and cover glass are 15 μm and 158 μm, respectively. Considering the group index (lithium niobate: 2.2, cover glass: 1.5) of the samples at the center frequency of the spectra, we estimate the thicknesses to be 7 μm for the lithium niobate thin film and 104 μm for the cover glass.

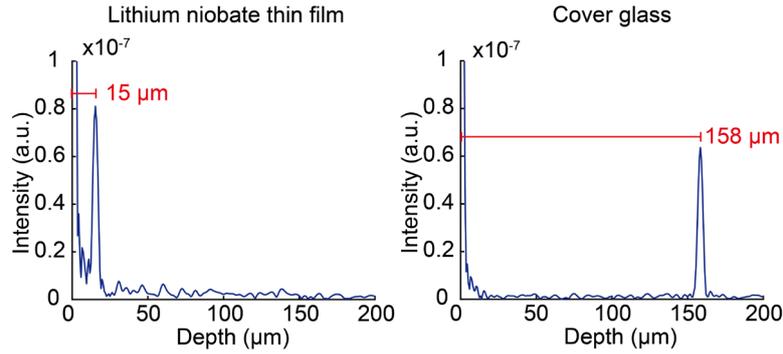

**Supplementary Figure 10**: Reference optical-depth measurements of the lithium niobate thin film and the cover glass used in the experiments.